\documentclass[galaxies,conferenceproceedings,submit,moreauthors,pdftex,10pt,a4paper]{mdpi} 

\firstpage{1} 
\articlenumber{}
\doinum{10.3390/------}
\pubvolume{xx}
\pubyear{2018}
\copyrightyear{2018}
\history{Received: date; Accepted: date; Published: date}

\pdfoutput=1

\usepackage{amssymb}
\usepackage{subfigure}
\usepackage{color}

\Title{Searching for Axion-Like Particles with X-ray Polarimeters}

\Author{Francesca Day $^{1,\dagger}$\orcidA{} and Sven Krippendorf $^{2,\dagger}$\orcidB{}}

\AuthorNames{Francesca Day and Sven Krippendorf}

\address{
$^{1}$ \quad DAMTP, University of Cambridge; francesca.day@maths.cam.ac.uk\\
$^{2}$ \quad Arnold Sommerfeld Center for Theoretical Physics, Ludwig-Maximilians-Universit{\"a}t M{\"u}nchen, Theresienstr.~37, 80333 M\"unchen, Germany}

\firstnote{These authors contributed equally to this work.}

\abstract{X-ray telescopes are an exceptional tool for searching for new fundamental physics. In particular, X-ray observations have already placed world-leading bounds on the interaction between photons and axion-like particles (ALPs). ALPs are hypothetical new ultra-light particles motivated by string theory models. They can also act as dark matter and dark energy, and provide a solution to the strong CP problem. In a background magnetic field, ALPs and photons may interconvert. This leads to energy dependent modulations in both the flux and polarisation of the spectra of point sources shining through large magnetic fields. The next generation of polarising X-ray telescopes will offer new detection possibilities for ALPs. Here we present techniques and projected bounds for searching for ALPs with X-ray polarimetry. We demonstrate that upcoming X-ray polarimetry missions have the potential to place world-leading bounds on ALPs.
}


\begin{document}

\section{Introduction}

The Standard Model (SM) of particle physics was completed by the discovery of the Higgs boson in 2012 \cite{1207.7214,1207.7235}, and reproduces a range of experimental results to astonishing accuracy \cite{PDG}. However, a host of observational and theoretical problems point to new physics as yet undiscovered. In particular, the discovery of dark matter \cite{Zwicky,Rubin,1502.01582} suggests that at least one fundamental particle remains to be found. Astrophysical observations represent a powerful tool in searching for new fundamental physics. From the discovery of dark matter, to powerful constraints on its self-interactions \cite{0704.0261} and its interactions with baryonic matter \cite{1503.02641,1704.03910}, astrophysics offers a potentially unique window to particle physics discovery. 

Axion-like particles (ALPs) are one of the leading candidates for new physics beyond the Standard Model. String theory models typically involve many ultra-light ALPs~\cite{hep-th/0602233,hep-th/0605206,1206.0819}. ALPs can act as both dark matter and dark energy \cite{1510.07633}, and the QCD axion offers an explanation for the non-observation of the neutron's electric dipole moment \cite{PQ}. ALPs interact weakly with SM particles. In particular, in the presence of a background magnetic field, ALPs and photons may interconvert \cite{Raffelt}. This effect has been used to search for ALPs with both astrophysics and laboratory experiments \cite{1403.5760}. X-ray astronomy is typically the optimal wavelength to search for low mass ($m_a \lesssim 10^{-12} \, {\rm eV}$) ALPs. For photons and ALPs propagating through typical astrophysical plasmas with electron densities $n_e \sim 10^{-4}  - 10^{-2} \, {\rm cm}^{-3}$, the X-ray spectrum in particular displays prominent oscillations, as described below. The effect of low mass ALPs at other wavelengths is often more subtle, and therefore X-ray astronomy has perhaps a crucial role to play in searching for axion-like particles. Existing X-ray data has been used to place world-leading bounds on the ALP-photon interaction by searching for characteristic modulations in the spectra of point sources passing through the magnetic fields of galaxy clusters \cite{1605.01043,1703.07354,1704.05256,1712.08313}.

The advent of polarizing X-ray telescopes is an exciting development in the quest for axion-like particles. As described below, the photon to ALP conversion probability depends on the direction of the photon's polarization with respect to the magnetic field direction. Therefore the existence of ALPs would lead to distinctive signatures in the X-ray polarization spectrum of sources passing through large magnetized regions. Upcoming missions such as IXPE \cite{IXPE} and Polstar \cite{1510.08358} open up new search strategies with the potential to discover ALPs. Such strategies have been discussed previously in \cite{1204.6187}, where photon to ALP conversion in the magnetic field of the Virgo supercluster is used. Under optimistic assumptions about this magnetic field, strong bounds on the ALP-photon coupling may be obtained.

We propose to search for signatures of ALPs in the X-ray spectrum of the linear polarization degree and polarization angle of point sources situated in or behind galaxy clusters. Clusters are known to host $\mathcal{O} (\mu {\rm G})$ magnetic fields over $\mathcal{O}({\rm Mpc})$ distances \cite{astro-ph/0410182}, making them ideal photon to ALP converters. This method has the potential to place world-leading bounds on the ALP-photon coupling, and therefore also has the potential to discover ALPs whose couplings are too weak to be detected by current experiments. In the next section, we will summarise the key results of ALP-photon physics. In section 3, we will describe our methodology for setting bounds on the ALP-photon coupling, and in section 4 we will present projected bounds from IXPE and Polstar data. In section 5 we will discuss these bounds and future prospects for ALP discovery with X-ray polarimetry.

\section{Axion-Like Particle Phenomenology}

The ALP-photon interaction is described by the Lagrangian

\begin{equation}
\mathcal{L} \supset g_{a \gamma \gamma}~a~\bf{E} \cdot \bf{B},
\end{equation}
where $a$ is the ALP field, $\bf{E}$ and $\bf{B}$ are the usual electric and magnetic fields and $g_{a \gamma \gamma}$ is the ALP-photon interaction strength. Assuming that the wavelength of the ALP and photon is much shorter than other scales in the environment, we can find a linearized equation of motion describing ALP-photon interconversion \cite{Raffelt}:

\begin{equation}
\label{propfree}
\left( \omega + \left( \begin{array}{ccc}
\Delta_{\gamma} & \Delta_{F} & \Delta_{\gamma a x} \\
\Delta_{F} & \Delta_{\gamma} & \Delta_{\gamma a y} \\
\Delta_{\gamma a x}  & \Delta_{\gamma a y} & \Delta_{a} \end{array} \right)
 - i\partial_{z} \right) \left( \begin{array}{c}
\mid \gamma_{x} \rangle\\
\mid \gamma_{y} \rangle\\
\mid a \rangle \end{array} \right) = 0,
\end{equation}
for a photon or ALP of energy $\omega$ travelling along the z direction. We have grouped the two photon polarizations $\mid \gamma_{x} \rangle$ and $\mid \gamma_{y} \rangle$ and the ALP $\mid a \rangle$ into a three component ALP-photon vector. The cluster's free electron density induces the photon mass term $\Delta_{\gamma} = \frac{-\omega_{pl}^{2}} {2 \omega}$, where $\omega_{pl} = \left( 4 \pi \alpha \frac{n_{e}}{m_{e}} \right) ^ {\frac{1}{2}}$ is the plasma frequency. The ALP mass term is $\Delta_{a} = \frac{-m_{a}^{2}}{\omega}$. In this work, we consider ultra-light ALPs where $m_a \lesssim \omega_{pl}$ ($m_a \lesssim 10^{-12}$ eV), and therefore approximate $\Delta_{a} = 0$. The mixing between the ALP and a photon polarized in the $i$ direction is given by the off-diagonal terms $\Delta_{\gamma a i} = \frac{g_{a \gamma \gamma} B_{i}}{2}$. Faraday rotation between the two photon polarizations is parametrized by $\Delta_{F}$. At X-ray energies, this effect is negligible, so we take $\Delta_{F} = 0$.

 The ALP-photon interconversion process is mathematically similar to neutrino oscillations, or to Faraday rotation with the ALP acting as a `third photon polarization'. In practice, we simulate ALP-photon interconversion by solving equation \eqref{propfree} numerically. In general, we find that the photon to ALP conversion probability $P_{\gamma \to a}$ is proportional to $B^2 g_{a \gamma \gamma}^2$, and to the total extent and coherence length of the magnetic field. Consistency with current observations requires $g_{a \gamma \gamma} \lesssim 5 \times 10^{-12} \, {\rm GeV}^{-1}$ \cite{astro-ph/9605197,astro-ph/9606028,1410.3747,1605.01043,1703.07354}.\\

\begin{figure}
\hfill
\subfigure[$P_{\gamma \to \gamma}$\label{oscillations:a}]{\includegraphics[scale=0.55]{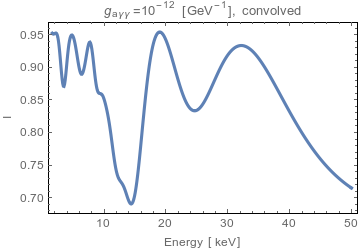}}
\subfigure[Q,U and V\label{oscillations:b}]{\includegraphics[scale=0.55]{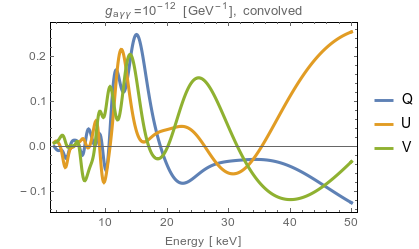}}
\centering
\subfigure[Linear polarization degree\label{oscillations:c}]{\includegraphics[scale=0.6]{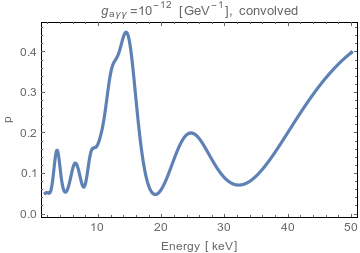}}
\caption{The photon survival probability (a), Stokes parameters (b) and degree of linear polarization (c) for initially unpolarized photons propagating from NGC1275 at the centre of Perseus, assuming the existence of ALPs with $g_{a \gamma \gamma} = 10^{-12} \, {\rm GeV}^{-1}$. The spectra are convolved with an energy resolution of 450 eV.}
\label{oscillations}
\end{figure}

ALP-photon interconversion leads to pseudo-sinusoidal oscillations in the spectrum of photons passing through a galaxy cluster. An example of such an oscillatory photon survival probability is shown in figure~\ref{oscillations:a}, for ALPs with $g_{a \gamma \gamma} = 10^{-12} \, {\rm GeV}^{-1}$. We have used the magnetic field experienced by photons from NGC1275, the AGN at centre of the Perseus galaxy cluster, described below.  The spectrum is convolved with an energy resolution of 450 eV, assuming a Gaussian energy dispersion. The energy resolution of the detector is crucial to ALP searches, as a poor energy resolution will smear out the characteristic oscillations. In this work, we assume that Polstar has an energy resolution of $\Delta E = 450$ eV below 10 keV, and $\Delta E = 4$ keV at 50 keV, rising linearly between 10 and 50 keV \cite{1510.08358}. IXPE has proportional counter energy resolution \cite{IXPE}. We will assume a $12 \%$ resolution at 5.9 keV, with $\Delta E \propto \sqrt{E}$ \cite{radiationbook}. The form of the photon survival probability spectrum depends on the (unknown) precise configuration of the cluster's magnetic field along the line of sight. However, the oscillatory structure is generic.

As well as the total flux, upcoming telescopes will measure the linear polarization and polarization angle of X-rays. As shown in equation \eqref{propfree}, only photons polarized parallel to the background magnetic field participate in ALP-photon interconversion. Therefore, the existence of ALPs induces an oscillatory linear polarization in the spectrum of an initially unpolarized source passing through a magnetic field. Even if the source is initially polarized, ALP induced oscillations in $p_{\rm lin}$ are seen. Figure~\ref{oscillations:b} shows the Stokes parameters and figure~\ref{oscillations:c} the degree of linear polarization for initially unpolarized light passing through Perseus' magnetic field. These are calculated from the ALP-photon vector in equation \eqref{propfree} as:

\begin{equation}
\label{stokes}
\begin{split}
I = |\gamma_x|^2 + |\gamma_y|^2~, \\
Q = |\gamma_x|^2 - |\gamma_y|^2~, \\
U = 2 {\rm Re} (\gamma_x \gamma_y^*)~, \\
V = -2 {\rm Im} (\gamma_x \gamma_y^*)~.
\end{split}
\end{equation}
To model light that is not initially fully polarised, we propagate both x and y polarised photons through the cluster magnetic field. We use initial ALP-photon vectors $(1, 0, 0)$ and $(0, 1, 0)$, leading to Stokes parameters $\{ S_x \}$ and  $\{ S_y \}$ respectively. Without loss of generality, we here take an initial polarisation angle $\psi = 0$. To find the measured Stokes parameters for an initial polarisation $p_0$ we take a weighted average of the Stokes parameters for $x$ and $y$ polarised beams as $S = \frac{1 - p_0}{2} S_x + \frac{1 + p_0}{2} S_y$. (Note that the normalisation of the Stokes parameters cancels in our final observables.) We can then find the linear polarisation degree $p_{\rm lin}$ and polarisation angle $\psi$ as:

\begin{equation}
\label{observables}
\begin{split}
p_{\rm lin} = \frac{\sqrt{Q^2 + U^2}}{I}~, \\
{\rm tan} ( 2 \psi ) = \frac{U}{Q}~.
\end{split}
\end{equation}



\section{Methodology}

Our first task is to simulate observations by IXPE and Polstar of NGC1275 both with and without the presence of ALPs. We assume that the flux of the source is given by a power law:

\begin{equation}
F(E) = 9 \times 10^{-3} \left( \frac{E}{\rm keV}\right)^{-1.8} {\rm cm}^{-2}{\rm s}^{-1}{\rm keV}^{-1},
\end{equation}
as found in \cite{1707.00176}.

We take a constant source spectrum for the degree of linear polarization $p_{\rm lin}$ and polarization angle $\psi$. NGC1275 is a type 1.5 AGN, so we do not expect to see significant energy dependence in its intrinsic polarization \cite{1709.03304}. Type 1.5 AGN show line ratios intermediate between those of type I and type II AGN \cite{Peterson}. In the unified AGN model, type 1.5 AGN are thought to have an inclination between type I and type II AGN of $45^{\circ} - 60^{\circ}$ \cite{1404.2417}. Future study of ALP searches with X-ray polarimetry should take into account detailed modelling of the source polarization and its uncertainties, as well as full instrumental simulation. For this preliminary study, we use a simplified source model and focus instead on the ALP physics. We consider the cases where NGC1275 has intrinsic linear polarization $p_{\rm lin} = 0$, $p_{\rm lin} = 1 \%$ and $p_{\rm lin} = 5 \%$. We must also consider the effect of the background (unpolarized) intracluster medium on the observations. The angular resolutions of IXPE and Polstar are $\sim 30'' $  and $\sim 60''$ respectively (half-power diameter) \cite{IXPE}. The signal to background ratio depends on the AGN luminosity, which varies on a timescale of years. Based on {\it Chandra} observations of NGC1275, we will conservatively assume a signal to background ratio of $R_{S/B} = 1/3$. As derived in \cite{1409.6214}, the optimal fraction of the total observation time spent observing the background rather than the source is $f_{\rm off} = \frac{\sqrt{1 + R_{S/B}} - 1}{R_{S/B}}$.\\

We bin the spectra such that each bin has an equal number of expected counts, and such that each bin width is greater than the telescope's energy resolution at the bin's central energy. We use 8 bins for IXPE and 10 for Polstar. When simulating measurements in each bin, we do not take any further account of telescope's energy resolution. This approximation is acceptable as our bins are wider than this resolution, but a more detailed study including full instrumental simulation should take this effect into account. Our procedure leads to bin widths that increase with energy. This ensures that we have sufficient statistics in each bin, and takes advantage of the fact that ALP induced oscillations are broader at higher energies. 

In the absence of detector simulation, we simply draw the measured $p_{\rm lin}$ and $\psi$ in each bin from the distribution derived in \cite{1409.6214}:

\begin{equation}
\label{measurement}
P(p_{\rm lin},\psi | p_0,\psi_0) = \frac{ \sqrt{I^2/W_2} p_{\rm lin} \mu^2}{2 \pi \sigma} \times {\rm exp} \left[ - \frac{\mu^2}{4 \sigma^2} \{p_0^2 + p_{\rm lin}^2 - 2 p_0 p_{\rm lin} {\rm cos}(2(\psi_0 - \psi)) - \frac{p_0^2 p_{\rm lin}^2 \mu^2}{2} {\rm sin}^2(2(\psi - \psi_0))\} \right],
\end{equation}
with 
\begin{equation}
W_2 = (R_S + R_{BG})T(1 - f_{\rm off}) + R_{BG} T f_{\rm off} \left(\frac{1 - f_{\rm off}}{f_{\rm off}}\right)^2,
\end{equation}
and 
\begin{equation}
\sigma = \sqrt{\frac{W_2}{I^2} \left( 1 - \frac{p_0^2 \mu^2}{2} \right)}.
\end{equation}
Equation \eqref{measurement} describes the probability of measuring a linear polarization degree $p_{\rm lin}$ and polarization angle $\psi$ in a given bin, given that the true values in that bin are respectively $p_0$ and $\psi_0$. $R_S$ and $R_B$ are the expected signal and background rates, $T$ is the total observation time and $I = R_S T (1 - f_{\rm off})$ is the total source intensity. Equation \eqref{measurement} assumes that the measured $p_{\rm lin}$ and $\psi$ are derived by reconstructing the Stokes parameters of the incident X-rays. It is assumed that the incoming photons are Poisson distributed, and neglects any systematic errors in the measurement. In future, a full detector simulation will yield more detailed and accurate projected bounds. However, we expect the discrepency between equation \eqref{measurement} and detector simulation to be subdominant to the uncertainty in the Perseus magnetic field discussed below. The quality of the polarization measurement depends crucially on the modulation factor $\mu$. $\mu$ is the amplitude of the modulation of scattered photons or photo-electrons for a completely polarized beam. We take $\mu$ as a function of energy from the values reported in \cite{IXPE} and \cite{1510.08358}. We also require the telescopes' effective area to compute $R_S$ and $R_B$. We assume a constant effective area of $750 \, {\rm cm}^2$ for IXPE \cite{IXPE} and take the functional dependence for Polstar given in~\cite{1510.08358}. We assume a total observation time $T = $ 1 Ms.\\

In the case without ALPs, we take a constant true linear polarization degree of $p_0 = 0$, $p_0 = 0.01$ or $p_0 = 0.05$, based on detailed modelling of AGN at various inclinations \cite{1309.1691,1709.03304}. Polarisation is generated when photons originating in the vicinity of the black hole scatter from the accretion disc. Our choice of relatively low intrinsic AGN polarizations makes our bounds conservative - for higher polarizations that are well above the minimum detectable polarization, oscillations in the polarization are easier to detect. We take an (arbitrary) constant true polarization angle $\psi_0 = 0$. In the case with ALPs, we use this constant linear polarization degree and angle as the initial state of a numerical simulation of equation \eqref{propfree}. In this way, we find the final state ALP-photon vector for X-rays originating from NGC1275 and propagating through the Perseus galaxy cluster. We then compute the final state Stokes parameters, and thus the final state $p_0$ and $\psi_0$. We require the electron density and magnetic field along the line of sight as inputs to \eqref{propfree}. Physically, these determine the effective mass difference between the ALP and the photon, and the strength of the ALP-photon mixing respectively. We take the electron density $n_e(r)$ of Perseus from \cite{astro-ph/0301482}. We take a central magnetic field strength of 25 $\mu$G, as inferred in \cite{astro-ph/0602622}. We assume the magnetic field falls off radially as $B \propto n_e^{0.7}$, based on detailed modelling of the Coma cluster \cite{1002.0594}. Information on the coherence length and power spectrum of Perseus' magnetic field is not available. We therefore choose values motivated by the structure of the magnetic field in the cool core cluster A2199 \cite{1201.4119}. We assume a minimum coherence length of 3.5 kpc, and a maximum coherence length of 10 kpc. We simulate the magnetic field along the line of sight from NGC1275 by drawing domain lengths randomly within this range with probability $P(l = x) \propto x^{-1.2}$. In each domain, the magnetic field takes a constant, random direction. We simulate a total length of 1 Mpc. These are the same underlying parameters as in~\cite{1707.00176}. In this way, we find the true linear polarization degree and angle in the presence of ALPs in each energy bin. As shown in figure \ref{oscillations}, the presence of ALPs induces an energy dependence of $p_0$ and $\psi_0$. We then randomly generate the measured $p_{\rm lin}$ and $\psi$ in each bin from equation \eqref{measurement}.\\

We now consider fitting a model without ALPs (i.e. a model with constant $p_0$ and $\psi_0$) to the measured (or, in our case, simulated) spectra. We do this by maximizing the likelihood given by equation \eqref{measurement}. We are now in a position to calculate projected bounds on ALPs from our simulated data. Our bounds are based on the fundamental principle that, if ALPs with a sufficiently strong coupling to photons exist, the constant model not including ALPs will be a bad fit to the polarimetry data. For each initial polarisation of the AGN, we simulate 10000 data sets \emph{without} ALPs. We then fit the constant no ALP model to each of these fake data sets, obtaining the likelihood value of each fit. For each initial polarisation, we then have a set of likelihoods $\{L_{\rm no ALP} \}$ for the fit of a no ALP model to data simulated without ALPs. In practice, we cannot know how good a fit the data actually obtained will be to standard astrophysical polarisation models. The strength of the bounds inferred from the real data will depend on the goodness of this fit. Here, we will take the average value of $\{L_{\rm no ALP} \}$, $L^{\rm av}_{\rm no ALP}$, to represent the goodness of fit when no ALPs are present in the data. We also consider a more conservative case in which the polarimetry data happen to be a rather bad fit to standard astrophysical models. In this case, we take the 90th percentile likelihood value, $L^{90}_{\rm no ALP}$,to represent the goodness of fit when no ALPs are present in the data.\\

If the effects of ALPs are sufficiently strong in the polarimetry data, a model not including  ALPs will be a bad fit to the data. We use this basic principle to place bounds on ALPs. For each $g_{a \gamma \gamma}$ value, we are considering two hypotheses:

\begin{itemize}
\item The null hypothesis $H_0$ that ALPs exist with coupling $g_{a \gamma \gamma}$ or higher to electromagnetism, and mass $m_a \lesssim 10^{-12}$ eV.
\item The alternative hypothesis $H_1$ that ALPs with coupling $g_{a \gamma \gamma}$ or higher to electromagnetism, and mass $m_a \lesssim 10^{-12}$ eV do not exist.
\end{itemize}

We seek to exclude the null hypothesis, and hence rule out ultra-light ALPs with electromagnetic coupling $g_{a \gamma \gamma}$ or higher. To do this we take a Bayesian approach, assuming a flat prior for the value of $g_{a \gamma \gamma}$. We compare $P({\rm data}|H_0)$ to $P({\rm data}|H_1)$ \cite{statsBook}. For each $g_{a \gamma \gamma}$ value, and each initial polarisation value, we use the following procedure:

\begin{enumerate}
\item Randomly generate 1000 different magnetic field realisations ${\bf B}_i$ for the line of sight to NGC1275.
\item For each ${\bf B}_i$, generate the ALP induced linear polarisation $p_0^i (E)$ and polarisation angle $\psi_0^i (E)$ spectra, by numerically propagating the initial photon vector through the cluster.
\item From each $\{ p_0^i(E), \psi_0^i(E) \}$ pair, generate 10 fake data sets by randomly sampling from equation \eqref{measurement}.
\item Fit the no ALP constant model to each of the resulting 10,000 fake data sets, and find the corresponding likelihoods $\{ L^{i}_{g a \gamma \gamma} \}$.
\item If fewer than $5 \%$ of the $\{ L^{i}_{g a \gamma \gamma} \}$ are equal to or higher than  $L^{\rm av}_{\rm no ALP}$ (or $L^{90}_{\rm no ALP}$ for the more pessimistic case), $g_{a \gamma \gamma}$ is excluded at the $95 \%$ confidence level.
\end{enumerate}
This method is equivalent to using $P({\rm data}|H_1)$ as a test statistic. In steps 2 - 4, we use Monte Carlo simulations to generate the null distribution of this test statistic. For comparison, we also computed bounds using the liklihood ratio test described in \cite{1603.06978}. This method leads to similar results, and is described in the appendix.

\section{Results}

\begin{table}[H]
\begin{tabular}{ l | l | l | l }
     & $0 \%$ & $1 \%$ & $5 \%$ \\
     \hline
  $L^{\rm av}_{\rm no ALP}$ & $1.2 \times 10^{-12} \, {\rm GeV}^{-1}$ & $1.2 \times 10^{-12} \, {\rm GeV}^{-1}$ & $6 \times 10^{-13} \, {\rm GeV}^{-1}$\\
  $L^{90}_{\rm no ALP}$ & $1.4 \times 10^{-12} \, {\rm GeV}^{-1}$ & $1.3 \times 10^{-12} \, {\rm GeV}^{-1}$ & $1.2 \times 10^{-12} \, {\rm GeV}^{-1}$ \\
\end{tabular}
\caption{\label{IXPEbounds}Projected upper limits on $g_{a \gamma \gamma}$ with IXPE. The columns correspond to different intrinsic polarisations of the AGN. The rows correspond to whether the average or 90th percentile likelihood value is used to characterize how well the no ALP model fits the simulated data.}
\end{table}

\begin{table}[H]
\begin{tabular}{ l | l | l | l }
     & $0 \%$ & $1 \%$ & $5 \%$ \\
     \hline
  $L^{\rm av}_{\rm no ALP}$ & $1.0 \times 10^{-12} \, {\rm GeV}^{-1}$ & $9 \times 10^{-13} \, {\rm GeV}^{-1}$ & $7 \times 10^{-13} \, {\rm GeV}^{-1}$\\
  $L^{90}_{\rm no ALP}$ & $1.2 \times 10^{-12} \, {\rm GeV}^{-1}$ & $1.2 \times 10^{-12} \, {\rm GeV}^{-1}$ & $1.2 \times 10^{-12} \, {\rm GeV}^{-1}$ \\
\end{tabular}
\caption{\label{Polstarbounds}Projected upper limits on $g_{a \gamma \gamma}$ with Polstar. The columns correspond to different intrinsic polarisations of the AGN. The rows correspond to whether the average or 90th percentile likelihood value is used to characterize how well the no ALP model fits the simulated data.}
\end{table}

\begin{figure}
\includegraphics[scale=1]{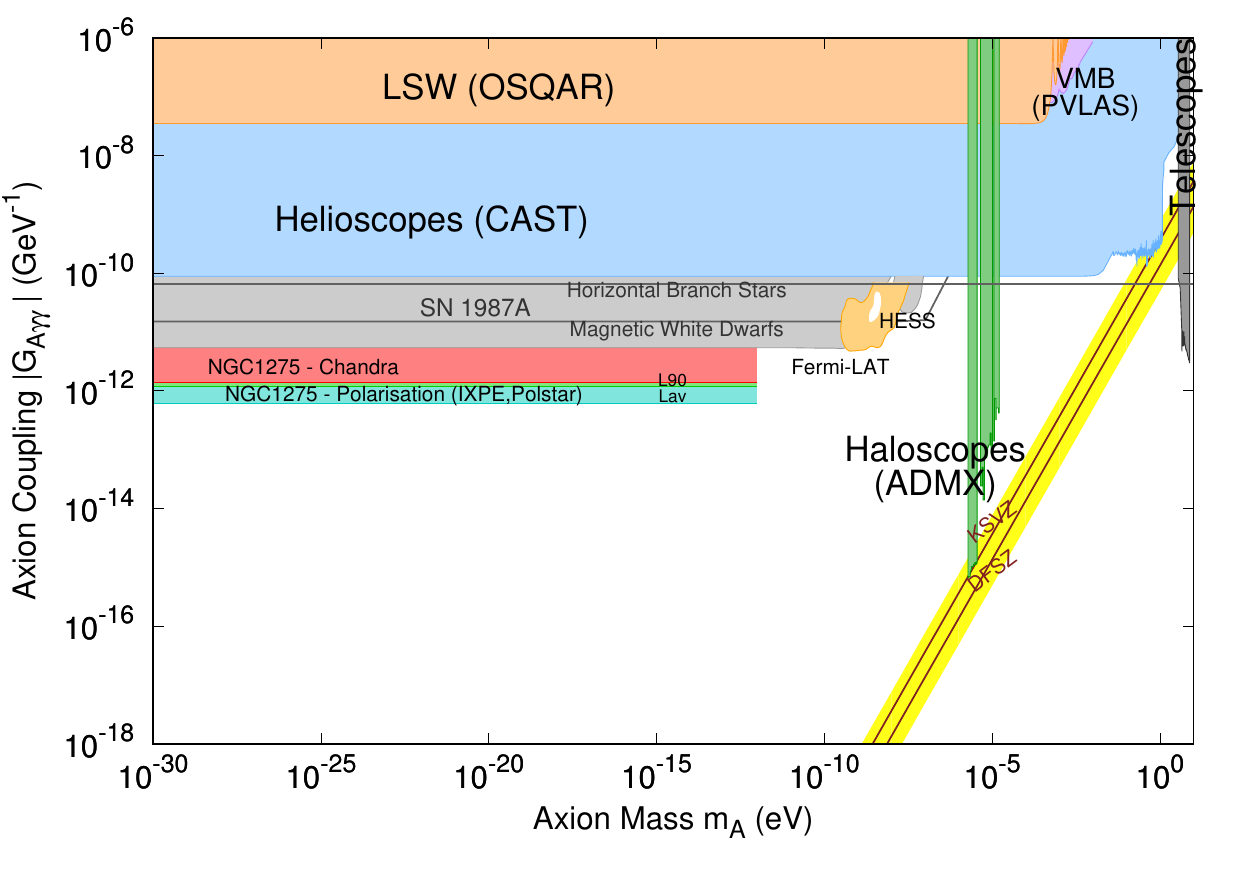}
\caption{Projected bounds on the ALP-photon coupling assuming a $5 \%$ initial polarization. Adapted from \cite{PDG}. We also include bounds set in \cite{1105.2083}. These are based on the non-observation of high degrees of linear polarization in magnetic white dwarfs, using simplified models of the strong magnetic fields these objects.}
\label{resultsPlot}
\end{figure}

Tables \ref{IXPEbounds} and \ref{Polstarbounds} show our projected bounds. These bounds are more constraining than existing bounds on low mass ALPs. For example, {\it Chandra} observations of NGC1275 give $g_{a \gamma \gamma} \lesssim 1.4 \times 10^{-12} \, {\rm GeV}^{-1}$ \cite{1605.01043}. A comparison of our projected bounds with current constraints is shown in figure \ref{resultsPlot}.

\section{Discussion}

Our preliminary projections of bounds from polarimetry measurements are approximately equal to or exceed by a factor of a few those obtained from X-ray flux only measurements of NGC1275 \cite{1605.01043}. These observations therefore have the potential to explore new regions of ALP parameter space. Even in the more pessimistic cases, obtaining similar bounds from different observables will increase our confidence in the exclusion of ALPs. Furthermore, improved bounds may be obtained by considering the polarisation and flux measurements together. For example, in the ALP case, regions of anomalously low flux are expected to be correlated with regions of anomalously high polarization.

One limitation of this preliminary study is that we approximate the source's polarization spectrum as a constant, rather than using state of the art astrophysical models and their uncertainties. When IXPE data is used to set bounds on ALPs, full instrumental and astrophysical modelling should be used. However, we note that ALPs give a distinctive oscillatory signature in the polarisation spectrum, which is unlikely to be mimicked by standard astrophysics. As shown above, our bounds are quite robust to the level of intrinsic polarization of the AGN, and to statistical fluctuations in the observations. We also mention in passing that ALPs can generate circular polarisation in an initially unpolarised spectrum \cite{1204.6187}. Although not accessible by upcoming X-ray polarimeters, this effect could be used in future to search for ALPs. A further source of uncertainty in many astrophysical bounds on ALPs, including those presented here, is our lack of knowledge of the astrophysical magnetic fields. If the true field in the Perseus galaxy cluster is significantly lower than assumed here, our bounds may be over-optimistic. In \cite{1605.01043}, Perseus' magnetic field is used to constrain ALP-photon conversion using data from {\it Chandra}. It was found that reducing the central magnetic field from $ 25 \mu {\rm G}$ to $ 10 \mu {\rm G}$ increased the upper bound on $g$ by a factor of $\sim 3$. We expect a similar uncertainty in our bounds to result from potential overestimation of Perseus' magnetic field.

X-ray polarimetry presents exciting prospects for fundamental physics, and in particular allows us to access novel signatures of ALPs. In this paper, we provide a proof of principle study of using oscillations in the linear polarization degree and angle of point sources shining through galaxy clusters to search for axion-like particles.

\acknowledgments{This work has been partially supported by STFC consolidated grant ST/P000681/1. FD is funded by Peterhouse, University of Cambridge. SK is funded by ERC Advanced Grant ''Strings and Gravity'' (Grant No. 320040). We are grateful to Joseph Conlon, Nicholas Jennings, Frederic Marin and MC David Marsh for illuminating discussions. We would also like to extend our thanks to Frederic Marin and the organisers of the Alsatian Workshop on X-ray Polarimetry, where many interesting discussions were had.}

\appendix
\section{Projected Bounds with a Likelihood Ratio Test}
We also find simulated bounds on ALPs from IXPE using the following method, adapted from~\cite{1603.06978}:
\begin{enumerate}
\item For each intrinsic source polarisation, simulate 1,000 data sets $\{ D_i \}$ with no ALPs present.

\item Simulate transfer matrices for each value of g considered and for 100 different magnetic field configurations $\{ B_j \}$. 

\item For each transfer matrix, find the final spectrum including ALPs for a range of different values for the intrinsic source polarisation degree $p_{\rm lin}^{\rm source}$ and angle $\psi^{\rm source}$. We take $p_{\rm lin}^{\rm source} = 0 - 10 \%$ in steps of $0.1 \%$ and $\psi^{\rm source} = 0 - \pi$ in steps of $\frac{\pi}{100}$, and we use an interpolating function derived from this data for the maximisation procedure later on.

\item We now fit the spectra with ALPs generated in the previous step to the fake data generated without ALPs. For each set $(g, B_j, D_i)$ we find the values of $p_{\rm lin}^{\rm source}$ and $\psi^{\rm source}$ that maximize the likelihood $L(g, B_j, p_{\rm lin}^{\rm source},\psi^{\rm source} | D_i) = {\displaystyle \prod_{\rm bins}} L_k(g, B_j, p_{\rm lin}^{\rm source},\psi^{\rm source} | D_i) $. In each bin $k$, $L_k$ is the probability of measuring the $p_{\rm lin}$ and $\psi$ values given by $D_i$, given that the true values are those predicted by an ALP model with parameters $(g, B_j, p_{\rm lin}^{\rm source},\psi^{\rm source})$. These are calculated from equation \eqref{measurement}. We thus obtain a set of maximised likelihoods $L(g,B_j | D_i)$.

\item For each value of $g$ and each $D_i$, sort the $L(g,B_j | D_i)$ obtained from different magnetic fields, and select the 95th quantile $L$ value, and the corresponding magnetic field. We thus obtain a set of likelihoods $L(g| D_i)$.

\item For each $D_i$, find the value of $g$, $\hat{g}$ that leads to the maximum $L(g| D_i)$.

\item We first consider the discovery potential of the data - i.e. the possibility of excluding a null hypothesis of no ALPs. For each $D_i$, we construct a test statistic $TS_i = -2 {\rm ln} \left( \frac{L(g=0|D_i)}{L(g = \hat{g}|D_i)} \right)$.

\item We have hence found the distribution of $TS$ under a null hypothesis of no ALPs. We find the threshold $TS$ value $TS_{\rm thresh}$ such that $95 \%$ of the $TS_i$ are lower than $TS_{\rm thresh}$. This value can be used to demonstrate our discovery potential for ALPs, by finding the $TS$ for some of our fake data with ALPs included. We note that this test statistic does not obey Wilk's theorem as our hypotheses are not nested.

\item We now turn to excluding values of $g$. Our null hypothesis is now that ALPs exist with some coupling $g$, and the alternative hypothesis $H_1$ is that $g \leq \hat{g}$. $H_1$ obviously includes the case where ALPs do not exist, but excluding ALPs with $g \leq \hat{g}$ should not be possible. Our test statistic for each $g$ is now $\lambda(g, D_i) = -2 {\rm ln} \left( \frac{L(g | D_i)}{L(\hat{g} | D_i)} \right)$.

\item We take the median value of $\lambda(g, D_i)$ over the $D_i$ to represent that $g$. So we now have simply $\lambda(g)$ for our test statistic.

\item We now need the null distribution of $\lambda(g)$ under the hypothesis that ALPs exist with coupling g. Following \cite{1603.06978}, we assume that $\lambda(g)$ and the test statistic for a null hypothesis of no ALPs, $TS$ above, have the same distribution, and therefore $\lambda(g)_{\rm thresh}$ = $TS_{\rm thresh}$. In \cite{1603.06978}, this assumption is tested with simulations for part of the parameter space. We therefore exclude a value of g if $\lambda(g) > TS_{\rm thresh}$.

\end{enumerate}

This method leads to the following bounds:

\begin{table}[H]
\begin{tabular}{ l | l | l | l }
     & $0 \%$ & $1 \%$ & $5 \%$ \\
     \hline
  $L^{\rm av}_{\rm no ALP}$ & $6 \times 10^{-13} \, {\rm GeV}^{-1}$ & $9 \times 10^{-13} \, {\rm GeV}^{-1}$ & $1.3 \times 10^{-12} \, {\rm GeV}^{-1}$
\end{tabular}
\caption{\label{IXPEboundsAppendix}Projected upper limits on $g_{a \gamma \gamma}$ with IXPE using the likelihood ratio method. The columns correspond to different intrinsic polarisations of the AGN.}
\end{table}

These bounds differ only by a factor of a few to those produced using the methodology above. One notable feature is that, using the likelihood ratio method, the bounds become less tight with increasing AGN polarisation, whereas using the method in section 3 this trend is reversed. There are two competing effects here: for higher AGN polarisations, the total polarisation degree will be higher and therefore oscillations in the polarisation will be easier to detect. Conversely, for higher AGN polarisations, ALP models are more likely to successfully fit Poisson fluctuations in the data, which is relevant in the method presented here.\\

We may also use this method to consider reconstructing an ALP signal (i.e. excluding the no ALP hypothesis) in the case where ALPs are present in the data. The focus of this paper is setting bounds, but we also demonstrate the ALP discovery potential of IXPE in the case of no intrinsic AGN polarisation. We proceed as follows:

\begin{enumerate}
\item For zero intrinsic source polarisation, simulate 10 data sets $\{ D_{i,g,B} \}$ for each $\{g, B \}$ pair, with $g$ running from $1 - 13 \times 10^{-13} \, {\rm GeV}^{-1}$ in steps of $1 \times 10^{-13} \, {\rm GeV}^{-1}$ and 5 different magnetic field configurations. We therefore have 50 fake data sets for each $g$.

\item Fit each $\{ D_{i,g,B} \}$ with spectra generated with ALPs of different $g$, as described in steps 2 - 7 above. (In this case we use 100 magnetic field configurations rather than 1000 in the interests of computational efficiency.) In this way we calculate a test statistic $TS_i$ for each $\{ D_{i,g,B} \}$. 

\item We compare $TS_i$ with the $TS_{\rm thresh}$ calculated in step 8 above. If $TS_i > TS_{\rm thresh}$ we may exclude the no ALP hypothesis at the $95 \%$ confidence level in that fake data set.

\item For each $g$, we find the proportional of the 50 corresponding fake data sets for each the no ALP hypothesis is excluded.

\end{enumerate}

We find that for $g \geq  1.1 \times 10^{-13} \, {\rm GeV}^{-1}$, the no ALP hypothesis can by excluded at the $95 \%$ confidence level in over $95 \%$ of the fake data sets. We therefore demonstrate that IXPE promises discovery as well as exclusion potential for ALPs.

\externalbibliography{yes}
\bibliography{axionproceedings.bib}

\end{document}